\documentclass[%
 reprint,
superscriptaddress,
showpacs,
 amsmath,amssymb,
 aps,
]{revtex4-1}

\usepackage{rotating}
\usepackage{stackengine}
\usepackage{graphicx}
\usepackage{color}
\usepackage{verbatim}
\usepackage{subfigure}
\usepackage{tikz}
\usepackage{tkz-graph}

\DeclareMathOperator{\sinc}{sinc}
\definecolor{Blue}{rgb}{0,0,1}
\definecolor{Red}{rgb}{1,0,0}
\definecolor{Black}{rgb}{0,0,0}
\definecolor{Blue2}{rgb}{0,0.4,1}
\newcommand{\db}[1]{{\color{Black} #1}}
\newcommand{\DBR}[1]{{\color{Black} #1}}
\newcommand{\DB}[1]{{\color{Black} #1}}
\newcommand{\D}[1]{{\color{Black} #1}}
\newcommand{\NBP}[1]{{\color{Black} #1}}

\begin{document}

\preprint{APS/123-QED}

\title{Spatio-spectral quantum networks in nonlinear photonic lattices
} 

\author{Natalia Costas} 
\affiliation{Galicia Supercomputing Center (CESGA), Avda.\ de Vigo S/N, Santiago de Compostela, 15705, Spain}
\author{Nadia Belabas} 
\affiliation{Centre de Nanosciences et de Nanotechnologies, CNRS, Universit$\acute{e}$ Paris-Saclay, 91120 Palaiseau, France}
\author{David Barral} 
\email{dbarral@cesga.es}
\affiliation{Galicia Supercomputing Center (CESGA), Avda.\ de Vigo S/N, Santiago de Compostela, 15705, Spain}

\begin{abstract}
Multiplexing information in different degrees of freedom is a natural solution to the scalability bottleneck in optical quantum communications and computing. However, for bulk-optics systems size, cost, stability, and reliability can make scalability either impractical or highly challenging to implement. We present a framework to engineer continuous \D{and discrete-variable} entanglement produced through nondegenerate spontaneous parametric down-conversion in $\chi^{(2)}$ nonlinear photonic lattices in spatial and spectral degrees of freedom that can alleviate the scalability challenge. We show how spatio-spectral pump shaping produce cluster states that are natively distributable in quantum communication networks and a resource for measurement-based quantum computing.
\end{abstract}

\date{\today}
\maketitle

\section{Introduction}

As the world becomes increasingly interconnected through technology, there is a growing need to enhance the capacity of communication channels. To support emerging technologies such as artificial intelligence, internet of things, and cloud computing, communication networks must \DB{handle} greater data volumes. In optical networks, when a channel reaches its capacity limit, multiplexing different degrees of freedom (DOF) becomes essential. Recent advancements in space-division multiplexing, combined with existing wavelength-division multiplexing, have enabled transmission rates exceeding few petabytes per second through a single optical fiber \cite{Puttnam2021}. 

Similarly, the future of quantum networks, which underpin emerging technologies like the quantum internet, distributed quantum sensing, and distributed quantum computing, faces a similar challenge \cite{Wehner2018, Zhang2021, Barral2024}. As the number of nodes in a quantum network or the complexity of quantum algorithms increases, so does the demand for more physical resources to process and transmit quantum information. Multiplexing DOF offers a practical solution to this challenge. In the optical domain, quantum information is encoded in space, path, frequency, time, polarization, and angular momentum, among others. \DB{Recent advances in multiplexing these DOF have shown remarkable progress in the quantum domain, demonstrating that quantum networks can scale more efficiently and handle growing demands like their classical counterparts \cite{Wright2022}.}

The transition from laboratory experiments to real-world applications is another critical aspect of quantum technologies. Small-footprint photonic circuits offer a promising solution, enabling the integration of various quantum operations—such as generation, processing, distribution, and measurement of quantum information—into a single, compact device \cite{Pelucchi2021, Moody2022}. These integrated optics solutions not only reduce the size but also improve stability, reliability, and scalability, making them ideal for building large, functional quantum networks \cite{Labonte2024, PsiQ2024, Aghaee2025}.

Quantum information may be encoded in variables that exhibit a continuous spectrum of eigenvalues, known as continuous variables (CV). In photonic systems, this encoding is realized through the fluctuations of field quadratures \cite{Braunstein2005}. Numerous tabletop experiments have successfully demonstrated CV quantum networks in the spatial, frequency, and temporal domains \cite{Asavanant2024}; and recently, integrated experiments in one DOF \cite{Lenzini2018, Nehra2022, Kouadou2023, Roman2024}. However, extending and adapting bulk-optics-based methods, such as sequential squeezing and entanglement, to systems involving a larger number of modes remains a highly demanding task. \D{Schemes based on multiplexing spatial and spectral DOF have been proposed for bulk-optics systems \cite{Yang2020, Barros2021}.} In this paper we show the potential of spatio-spectral multiplexed encoding of quantum information in the propagating modes generated in a nonlinear $\chi^{(2)}$ photonic lattice without bulk-optics analogous. The distributed simultaneous nonlinearity and evanescent coupling configuration of the photonic lattice together with the spectral properties of both the waveguides and the interacting fields parallelize multimode spatio-spectral transformations. Previous works have analyzed the potential of nonlinear photonic lattices in discrete variables \cite{Solnstev2012, Kruse2013, Belsley2020, Hamilton2022, He2024, Delgado2024}. We propose a general framework for nonlinear waveguide arrays in CV \DB{--that includes discrete variables (DV) as a limit case--} that allows engineering two and three dimensional \DB{graph and} cluster states naturally distributable in quantum networks and \D{that are} a resource for quantum computing. 

\begin{figure*}[t]
    \includegraphics[width=0.85\textwidth]{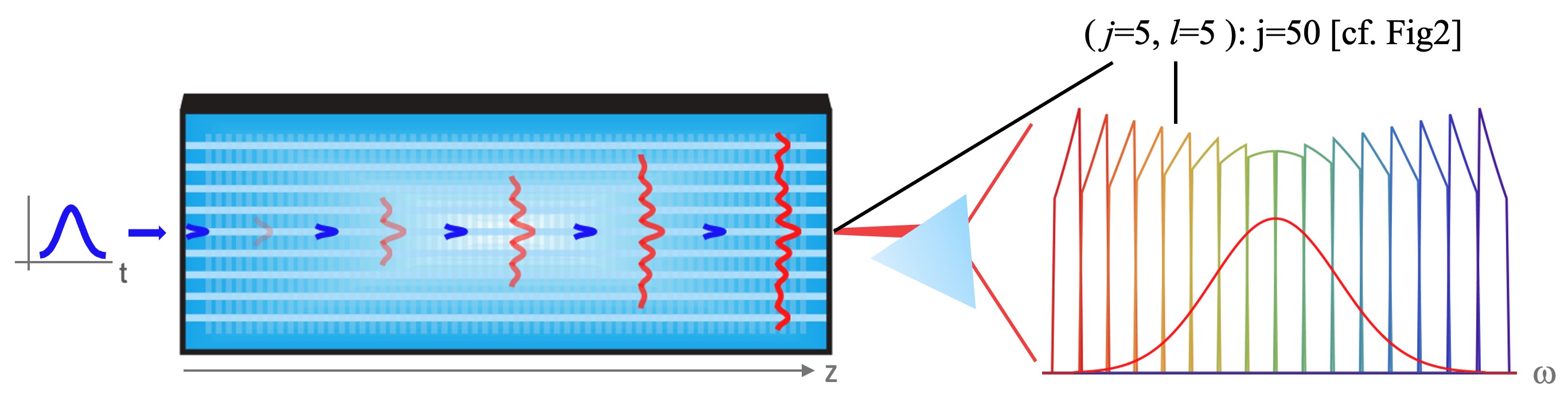}
\caption{\label{F1}\small{Sketch of a nonlinear photonic lattice and measurement basis in a given spatio-spectral mode (\DBR{fixel $\mathsf{j}$}) associated to a given waveguide \DBR{(pixel $j$)} and frequency band \DBR{(frexel $l$)}. Left: a pump pulse (blue) is coupled to the center waveguide of a $N=9$ lattice producing SPDC (red) that spreads accordingly to a coupling profile. Right: orthonormal homodyne measurement basis --frexels (rainbow)-- for a Gaussian-shaped local oscillator \NBP{(decomposed in frequency bins, normalized in power --rainbow slices)}, and, \NBP{in red, the} signal marginal of the joint spatio-spectral amplitude (JSSA) (equally for idler) projected on the frexel basis. \D{This example shows the measurement of 16 frexels corresponding to one pixel, hence $9\times16=144$ fixels. In general the system gives access to $N\times L=\mathsf{N}$ fixels with the notations of Table I.}}}
\end{figure*}

\section{Dynamics of a nonlinear photonic lattice}

A nonlinear photonic lattice consists of $N$ identical $\chi^{(2)}$ waveguides in which spontaneous downconversion (SPDC) and nearest-neighbor evanescent coupling between the generated fields take place (see Fig. \ref{F1}) \cite{Christo2003, Solnstev2014, Raymond2024}. We consider type 0/I downconversion where in each waveguide an input harmonic field at frequency $\omega_{h}$ is downconverted into signal (s) and idler (i) fields respectively at frequency $\omega_s$ and $\omega_i =\omega_{h}-\omega_{s}$ with identical polarization. Other strategies as type II downconversion can be equally implemented adding polarization as an extra DOF \cite{Victor2021}. The buildup of the nonlinear interaction is driven by the ability of propagating the interacting waves with the same velocity or wave-vector phase-matching. Birefringence is not always applicable, or not to the highest second-order tensor component. A possible solution is the periodic modulation of the nonlinear coefficient to \DB{quasi-phasematch} the propagation constants: $\Delta\beta(\omega_{s},\omega_{i})\equiv\beta(\omega_{h})-\beta(\omega_s)-\beta(\omega_{i})-2\pi/\Lambda=0$, with $\beta$ the propagation constant and $\Lambda$ the poling period. For instance in lithium niobate, where birefringence does not give access to its highest second-order tensor component $d_{33}$, periodic poling enables access to it. State-of-the-art poled waveguides have demonstrated up to 8 dB of squeezing \cite{Kashiwazaki2023}. Below, we consider that phase-matching is just produced \DB{in the region with evanescent coupling between generated modes \cite{Barral2021OE}. The energy of the signal/idler} modes propagating in each waveguide is exchanged between the coupled waveguides through evanescent waves, whereas the interplay of the second harmonic waves is negligible for the considered propagation lengths due to their high confinement into the guiding region. We set our calculation in the regime of pump undepletion.

\begin{table*}[t]
\centering
\begin{tabular}{c c c c c}
\hline\hline
 &\vline  &  \qquad Spatial (guided) \qquad   &  \qquad Spectral \qquad  &  \qquad Spatio-spectral \qquad \\[0.5ex]
\hline
Individual modes &\vline & \qquad Pixel: $\hat{\mathcal{A}}_{j}$ & \qquad Frexel: $\hat{\mathcal{A}}^{l}$ & \qquad Fixel: $\hat{\mathcal{A}}_{j}^{l} \equiv \hat{\mathcal{A}}_{\mathsf{j}}$ \\
\hline
Linear supermodes &\vline & \qquad L-pixel: $\hat{\mathcal{B}}_{k}$  & \qquad n.a. & \qquad  L-fixel: $\hat{\mathcal{B}}_{k}^{l} \equiv \hat{\mathcal{B}}_{\mathsf{k}}$ \\
\hline
Nonlinear supermodes &\vline & \qquad*\cite{Barral2020b} & \qquad*\cite{Roslund2014} & \qquad N-fixel: $\hat{\mathcal{C
}}_{\mathsf{m}}$  \\
\hline
Number of modes &\vline & \qquad $N$ & \qquad $L$ & \qquad $\mathsf{N}=N\times L$ \\
\hline
\end{tabular}
\caption{\label{T1}\small{Notation used to designate the mode bases and the annihilation operators related to spatial, spectral and spatio-spectral individual modes, linear supermodes and nonlinear supermodes. The number of modes per DOF is \D{given in the last line. The elements with * are not addressed in this article \NBP{but detailed in the mentioned refs \cite{Barral2020b,Roslund2014}}. n.a. stands for not applicable.}}}
\end{table*}

\begin{figure*}[t]
    \includegraphics[width=0.95\textwidth]{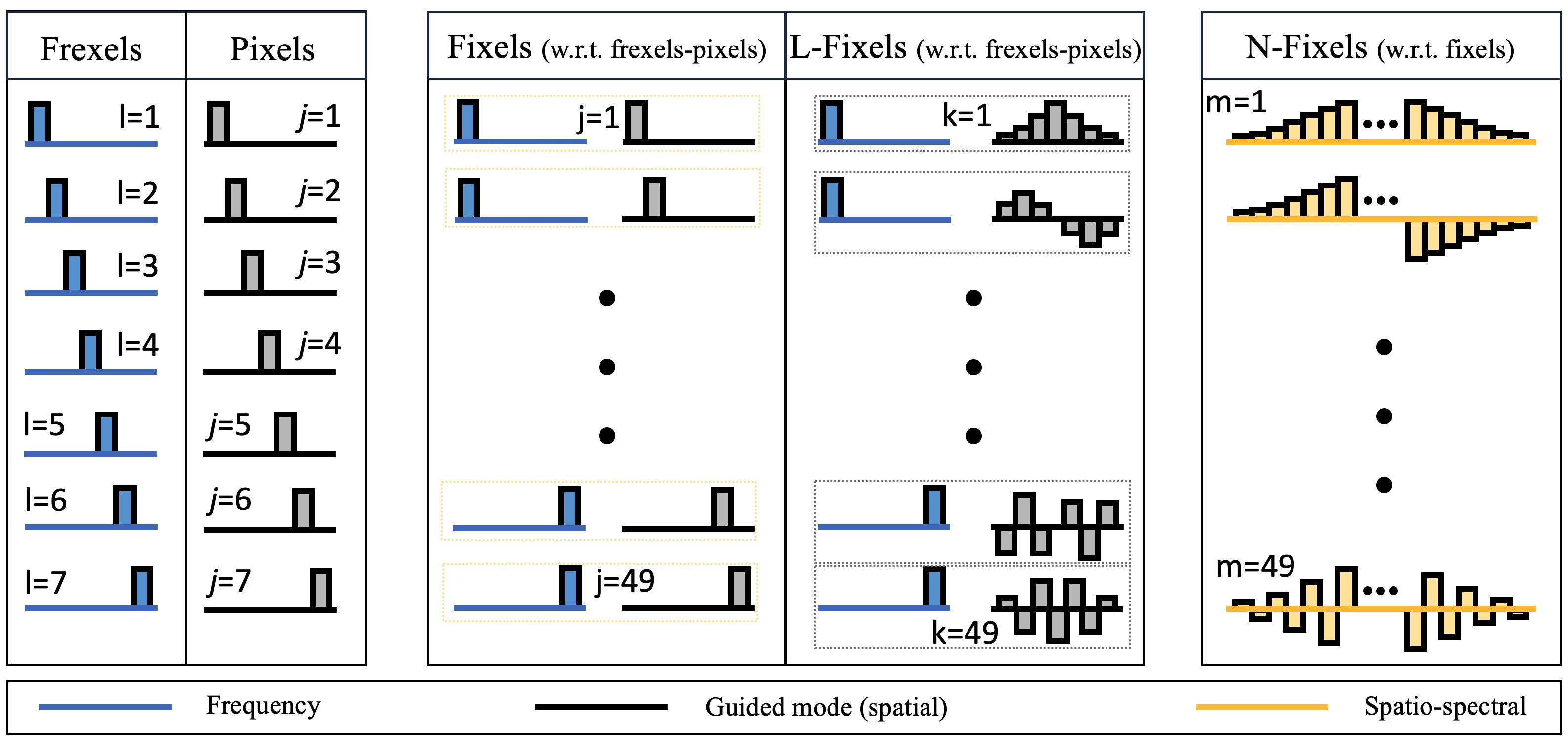}
\caption{\label{F2}\small{\NBP{Sketch of mode bases appearing in the text. In the left we display spectral and spatial non-overlapping (individual) mode basis: frexels and pixels, respectively. On the center we show two families of spatio-spectral modes: fixels and L-fixels. Fixels are non-overlapping modes separable in spectral and spatial DOF. L-fixels are non-overlapping modes in frequency and overlapping modes (supermodes) in space, separable in spectral and spatial DOF. The dotted boxes represent single spatio-spectral modes (yellow for fixels, black for L-fixels). On the right we introduce a third family of spatio-spectral modes: N-fixels. They are overlapping modes in frequency and space, in general inseparable in spectral and spatial DOF, thus only describable in terms of fixels (or L-fixels). This basis is not accessible experimentally unless spectral and spatial DOF are decoupled. Fixels and L-fixels are independent of $z$ (but L-fixels depend on the coupling profile), while N-fixels depend on coupling and pump profiles, and on $z$. The nonlinear photonic lattice produces quantum correlations in fixels and L-fixel bases, and independent squeezing in the N-fixel basis. In the figure we have set as an example $L=7$ spectral and $N=7$ spatial modes --thus $\mathsf{N}=49$ spatio-spectral modes. Blue, black and yellow abscissas represent frequency, guided mode, and fixels, respectively. Rectangles represent relative amplitudes with respect to other basis. w.r.t. stands for with respect to.}}}
\end{figure*}

The physical processes taking place in $\chi^{(2)}$ waveguides can be described by a dynamical operator $\hat{\mathcal{M}}$ obtained quantizing the flux of momentum of the electromagnetic fields \cite{Horoshko2022}. Particularly, we tackle the problem of broadband downconversion with an arbitrary pump, i.e. we assume that the \DB{downconverted spectrum $\Delta_{\text{SPDC}}$ is much broader than the pump bandwidth such that we can neglect the dependence of the nonlinear susceptibility $\chi^{(2)}$} on frequency \cite{Dayan2007}. Thus, the following Heisenberg equation is obtained for an array of $N$ evanescently coupled nonlinear waveguides with an arbitrary pump waveform in the SPDC regime
\begin{align}\nonumber
&\frac{d \hat{{A}}_{j}(\omega_s ,z)}{d z}=\\ \nonumber
& i C_{M}[f_{j-1}(\omega_s) \hat{{A}}_{j-1}(\omega_s,z)+ f_{j}(\omega_s)\hat{{A}}_{j+1}(\omega_s ,z)] \\  \label{one}
&+ i g \int \alpha_{j}(\omega_{h}) \,e^{i\Delta\beta(\omega_{s},\omega_{i}) z} \hat{{A}}_{j}^{\dagger} (\omega_i ,z) d\omega_{i},
\end{align}
where $\hat{{A}}_{0}=0$ and $\hat{{A}}_{N+1}=0$, $f_{0}=f_{N}=0$ and $j=1,\dots, N$ is the individual mode index \db{corresponding to each waveguide}. \DB{$C_{M}$ is the largest linear coupling strength at the waveguide-array design operating frequency $\omega_{h}/2$, and $f_{j}(\omega)$ are the elements of the coupling profile $\vec{f}$, and $z$ is the coordinate along the direction of propagation}. $\hat{{A}}_{j}(\omega_s ,z) \equiv \hat{{A}}_{j}^{s}$ and $\hat{{A}}_{j}(\omega_i ,z)\equiv \hat{{A}}_{j}^{i}$ are monochromatic slowly-varying amplitude annihilation operators of signal and idler photons corresponding to the $j$th waveguide \DB{--that we denote as pixel basis (Table \ref{T1} \DBR{and Figure \ref {F2}} summarize the different bases introduced in the paper)}-- fulfilling \db{local} commutation relations $[\hat{{A}}_{j}(\omega,z), \hat{{A}}_{j'}^{\dagger}(\omega',z)]=\delta(\omega-\omega')\delta_{j,j'}$. We consider the nonlinear coupling constant $g$ --proportional to $\chi^{(2)}$ and to the spatial overlap of the signal and harmonic fields in each waveguide-- equal for all waveguides. We work at the degeneracy point of the phase matching $\omega_{h}=2\omega_s$. $\alpha_{j}(\omega_{h})\equiv \langle \alpha \vert \hat{{A}}_{j}(\omega_{h}) \vert \alpha \rangle=\sqrt{P_{h}}\,\eta_{j}\,\Omega(\omega_{h}=\omega_{s}+\omega_{i})$ is the complex amplitude of a strong coherent undepleted pump field propagating in the $j$th waveguide, where $P_{h}$ is the total pump power in the array, $\eta_{j}$ is the normalized complex amplitude of pump power directed to each \db{waveguide} $j$, and $\Omega$ is the normalized spectral pump complex amplitude that feeds the production of a pair of photons in frequencies $\omega_s$ and $\omega_i$. \DB{The spatial and spectral pump profiles can be tuned by spatial and temporal pulse shapers that independently manipulate the relative phases and amplitudes of propagating waveguide modes and frequency modes, respectively \cite{Barral2020, Michel2021}}. Experimentally, distributable frequency modes --or frexels (\DBR{see Fig. \ref{F2})}-- are measured in bands of a given bandwidth by either shaping a local oscillator (LO) or in a multipixel homodyne detector \cite{Roslund2014, Arzani2018,Fabre2020}. These modes can be spatially separated by means of a dispersive element such as a grating or a prism and a microlens array, or fiber-based wavelength-division multiplexing \cite{Michel2021, Liu2016}. The orthogonality of these modes is guaranteed as they do not overlap \DB{spectrally}. We thus discretize the downconverted spectrum in $L$ bands centered at frequencies $\omega_{l}$ such that each signal(idler) frexel mode is labelled with $l(l')$. The frexel-pixel operators $\hat{\mathcal{A}}_{j}^{l}(z)$ \DB{--the fixel basis--} are related to the monochromatic frequency-mode operators $\hat{A}_{j}(\omega, z)$ simply by a basis transformation between the discrete and the continuous basis $\hat{\mathcal{A}}_{j}^{l}(z)=\int d\omega\, \xi^{l}(\omega) \hat{A}_{j}(\omega, z)$, fulfilling local commutation relations $[\hat{\mathcal{A}}_{j}^{l}(z), \hat{\mathcal{A}}_{j'}^{ l' \dagger}(z)]=\delta_{j,j'}\delta_{l,l'}$, where $\xi^{l}(\omega)$ corresponds to normalized frexel modes of width $\Delta_F$ taken as real for simplicity. The extent and number of frexel modes are \DB{fixed by} experimental constraints: they result from i) the bandwidth of the local oscillator and from ii) the resolution of a LO pulse shaper or a multipixel detector, the channel bandwidth in a dense wavelength division multiplexer (DWDM), etc \DBR{\cite{Michel2021}}. 

Considering coupling only between nearest-neighbor waveguides, a linear waveguide array --Equation \ref{one} with $g=0$-- presents spatial linear supermodes \DB{(L-pixels)} $\hat{\mathcal{B}}_{k}$, i.e. propagation eigenvectors \cite{Kapon1984}. These eigenvectors form a basis and are represented by an orthogonal matrix {${\bf M}\equiv {\bf M}(\vec{f})$} with real elements $M_{k,j}$. The L-pixels are the same for signal and idler frequencies as long as the coupling profile \DB{is constant in the considered bandwidth $\vec{f}(\omega) \approx \vec{f}$}. We consider a SPDC bandwidth $\Delta_{\text{SPDC}}$ where this condition holds and discuss the validity of this assumption in the Appendix. \DB{Pixels and L-pixels} are thus related by $\hat{\mathcal{B}}_{k}^{l}=\sum_{j=1}^{N} M_{k,j}\,\hat{\mathcal{A}}_{j}^{l}$. These supermodes are orthonormal $\sum_{j=1}^{N}M_{k,j} M_{k',j}= \delta_{k,k'}$, with a spectrum of eigenvalues $\lambda_{k}\equiv\lambda_{k}(C_{M},\vec{f})$. $\lambda_{k}$ is the propagation constant of the $k$th \DB{L-pixel}. Recent developments have broken the \D{static} structure of integrated optical lattices enabling the modification of coupling constant profiles on demand \cite{Yang2024, Onodera2024}.

To simplify the analysis we order the modes in $L$ blocks composed of $N$ spatial modes in given frequency bands centered at  $\omega_{l}$ with the following indices respectively for the fixels and linear spatio-spectral supermodes \DB{(L-fixels)}
\begin{equation}\nonumber
\mathsf{j}=j+(l-1)N, \quad \mathsf{k}=k+(l-1)N, 
\end{equation}
with $\{\mathsf{j}, \mathsf{k}\}=1, \dots, \mathsf{N}$ and $\mathsf{N}\equiv N\times L$ \cite{Note1}. The indices are ordered using $l$: for $l=1$ we have $\mathsf{j(k)}=1, \dots, N$, for $l=2$ we have $\mathsf{j(k)}=N+1, \dots, 2N$, etc. The \DB{L-fixel} transformation is then
\begin{equation}\nonumber
\hat{\mathcal{B}}_{\mathsf{k}}=\sum_{\mathsf{j}=1}^{\mathsf{N}} \mathsf{M}_{\mathsf{k}, \mathsf{j}}\,\hat{\mathcal{A}}_{\mathsf{j}},
\end{equation}
with $ {\mathsf{M}}$ a block diagonal matrix with elements $ \mathsf{M}_{\mathsf{k}, \mathsf{j}}\equiv{M}_{{k},{j}}$. Using slowly-varying \DB{L-fixel} amplitudes $\hat{B}_{\mathsf{k}}=\hat{\mathcal{B}}_{\mathsf{k}}\,e^{-i\lambda_{\mathsf{k}} z}$ the following propagation equation is obtained straightforwardly
\begin{equation}\label{Hei}
\frac{d\hat{B}_{\mathsf{k}}}{d z}= i g\,\sqrt{P_{h}}\sum_{\mathsf{k}'=1}^{\mathsf{N}} \mathcal{L}_{\mathsf{k},\mathsf{k}'}(z) \hat{B}_{\mathsf{k}'}^{\dag},
\end{equation}
where $\mathsf{k}$ and $\mathsf{k'}$ are two \DB{L-fixels} coupled by a function ${\mathcal{L}}(z)$ with elements given by $\mathcal{L}_{\mathsf{k},\mathsf{k}'} (z)= \tilde{\mathcal{L}}_{k,k'} (z) \tilde{\mathcal{L}}^{l,l'} (z)$, where
\begin{align} \label{JSA1}
&\tilde{\mathcal{L}}_{k,k'} (z)=\sum_{j=1}^{N} M_{k,j} M_{k',j} \eta_{j}\,e^{-i(\lambda_k + \lambda_{k'}) z},  \\  \label{JSA2}
&\tilde{\mathcal{L}}^{l,l'} (z)=\iint d\omega_s d\omega_i \xi^{l}(\omega_s) \xi^{l'}(\omega_i) \Omega(\omega_{s}+\omega_{i}) e^{i\Delta\beta(\omega_s, \omega_i) z}.
\end{align}
$\mathcal{L}(z)$ is a complex matrix which gathers all the information about the spatio-spectral shape of the pump and the phasematching. The symmetric complex matrices $\tilde{\mathcal{L}}_{k,k'}$ and $\tilde{\mathcal{L}}^{l,l'}$ couple respectively \DB{L-pixels} and frexels and, as we show below \D{in Eq. \eqref{intL}}, propagation couples both. Note the outstanding symmetry between the two expressions with a change of basis, a pump function and a phasematching function.   

Remarkably, unlike broadband frequency modes, frexels preserve local multiplication. If the resolution of the frexel basis $\Delta_F^{-1}$ is large enough, the frequency-dependent functions are approximately constant within each frexel, and we can approximate $\Omega(\omega_{s}+\omega_{i})$ and $\Delta\beta(\omega_s, \omega_i)$ by their frexel versions
\begin{align} \label{apA}
\Omega^{l,l'}&=\iint d\omega_s d\omega_i \xi^{l}(\omega_s) \xi^{l'}(\omega_i) \Omega(\omega_{s}+\omega_{i}),\\ \label{apB}
\Delta\beta^{l,l'}&=\iint d\omega_s d\omega_i \xi^{l}(\omega_s) \xi^{l'}(\omega_i) \Delta\beta(\omega_{s},\omega_{i}),
\end{align}
such that $\tilde{\mathcal{L}}^{l,l'} (z) \approx \Omega^{l,l'} e^{i\Delta\beta^{l,l'}z}$. This coarse-grained description of the spectral functions can be used to get insight about the dynamics of the full system although analytical calculations rigorously hold only
for \eqref{JSA2}. 

The symmetries of the \DB{L-fixels} enable solving Eq. \eqref{Hei} analytically for any gain regime with suitable pump profiles \cite{Barral2020, Barral2020b,Barral2021PRR}. However, in the general case, we can solve it in the low gain regime where space-ordering effects can be neglected \cite{Christ2013}. Hence, the formal solution to Equation (\ref{Hei}) at given $z$ is written as
\begin{equation} \label{Bsol}
\begin{small}
\begin{pmatrix}
\vec{B}(z) \\
\vec{B}^{\dag}(z)
\end{pmatrix}= \exp \left\lbrace 
\Gamma(z)
\begin{pmatrix}
0 & \mathsf{f}(z) \\
\mathsf{f}^{*}(z) & 0
\end{pmatrix}
  \right\rbrace
  \begin{pmatrix}
\vec{B}(0) \\
\vec{B}^{\dag}(0)
\end{pmatrix},
\end{small}
\end{equation}
with $\vec{B}\equiv$ $(\hat{B}_{\mathsf{1}}, \dots, \hat{B}_{\mathsf{k}}, \dots, \hat{B}_{\mathsf{N}})^{T}$, $\Gamma(z)=g \sqrt{P_{h}} z$ the total nonlinear amplitude and $\mathsf{f}(z)$ the normalized joint spatio-spectral amplitude (JSSA), with elements given by
\begin{align} \nonumber
&\mathsf{f}_{\mathsf{k},\mathsf{k}'}(z)= \eta_{k,k'} \iint d\omega_s d\omega_i \xi^{l}(\omega_s) \xi^{l'}(\omega_i) \Omega(\omega_{s}+\omega_{i})\\ \label{intL}
&\times \sinc(\frac{\Delta{\beta}(\omega_s,\omega_i)-(\lambda_k + \lambda_{k'})}{2}z) \,e^{i\frac{\Delta{\beta}(\omega_s,\omega_i)-(\lambda_k + \lambda_{k'}) }{2}z},
\end{align}
with $\eta_{k,k'}=\sum_{j=1}^{N} M_{k,j} M_{k',j} \eta_{j}$. 
$\mathsf{f}_{\mathsf{k},\mathsf{k}'}(z)$ couples pairs of \DB{L-fixels} $\mathsf{k}$ and $\mathsf{k}'$ and, in particular, the $\sinc$ function couples spatial and spectral DOF. Thus, the spectral features of the waveguides, the evanescent coupling profile as well as the spatial and spectral shape of the pump enable engineering the JSSA. This matrix contains all information about individual-mode correlations in the DV regime ($\Gamma(z) <<$) through a simple change of basis in the spatial DOF \cite{Delgado2024}. 

Under the high resolution approximation for frexels of Eqs. \eqref{apA}-\eqref{apB}, Eq. \eqref{intL} is separable into spatio-spectral pump and phasematching functions as
\begin{align} \label{intL2}
\mathsf{f}_{\mathsf{k},\mathsf{k}'}(z) \approx \alpha_{\mathsf{k},\mathsf{k}'} \Phi_{\mathsf{k},\mathsf{k}'}(z),
\end{align}
with $\alpha_{\mathsf{k},\mathsf{k}'}=\eta_{k,k'} \Omega^{l,l'}$, $\Phi_{\mathsf{k},\mathsf{k}'}(z)=\sinc(\frac{\Delta\tilde{\beta}_{\mathsf{k},\mathsf{k}'} z}{2}) \,e^{\frac{i \Delta\tilde{\beta}_{\mathsf{k},\mathsf{k}'} z}{2}}$, and $\Delta\tilde{\beta}_{\mathsf{k},\mathsf{k}'}=\Delta\beta^{l,l'}-(\lambda_k + \lambda_{k'})$.

The JSSA of Eqs \eqref{intL} and \eqref{intL2} is symmetric, i.e. it is invariant under the change of indices $\mathsf{k} \leftrightarrow \mathsf{k}'$. Using this property, the solution to Equation (\ref{Bsol}) can be simplified through a full nonlinear supermode \DB{(N-fixel)} basis $\hat{\mathcal{C}}$, given by $ \hat{\mathcal{C}}_{\mathsf{m}}=\sum_{\mathsf{k}=1}^{\mathsf{N}}\Upsilon_{\mathsf{m},\mathsf{k}}^{\dag}(z)\,\hat{{B}}_{\mathsf{k}}$, where $\Upsilon(z)$ is an unitary matrix which diagonalizes $\mathsf{f}(z)$ by a congruence transformation -- the Autonne-Takagi transformation-- obtaining  a real diagonal matrix with non-negative entries $\Lambda(z)$ \cite{Cariolaro2016}. Equation (\ref{Bsol}) in terms of \DB{N-fixels} is thus simply given by 
\begin{equation}\label{Csol}
\hat{\mathcal{C}}_{\mathsf{m}}(z)=\cosh[{r}_{\mathsf{m}}(z)] \,\hat{\mathcal{C}}_{\mathsf{m}}(0)+\sinh[{r}_{\mathsf{m}}(z)]\,\hat{\mathcal{C}}_{\mathsf{m}}^{\dag}(0), 
\end{equation}
with $m=1,\dots, \mathsf{N}$, and where ${r}_{\mathsf{m}}(z)=\Gamma(z)\Lambda_{\mathsf{m},\mathsf{m}}(z)$ are the downconversion gains at a propagation distance $z$. Each spatio-spectral \DB{N-fixel} thus appears as a broadband non-local single-mode squeezed state. 

In terms of \DB{fixels}, the solution to the nonlinear system is
\begin{align}\label{IndGenSol}
\hat{\mathcal{A}}_{\mathsf{j}}(z)=\sum_{\mathsf{j'}} U_{\mathsf{j},\mathsf{j}'}(z) \hat{\mathcal{A}}_{\mathsf{j}'}(0)+V_{\mathsf{j},\mathsf{j}'}(z)\hat{\mathcal{A}}_{\mathsf{j}'}^{\dag}(0),
\end{align}
with
\begin{align}\nonumber
U_{\mathsf{j},\mathsf{j}'}(z)&=\sum_{\mathsf{k, m}}\mathsf{M}_{\mathsf{j},\mathsf{k}} \Upsilon_{\mathsf{k},\mathsf{m}}(z) \mathsf{M}_{\mathsf{m},\mathsf{j}'}\, e^{i \lambda_{\mathsf{k}} z} \cosh[{r}_{\mathsf{m}}(z)], \\  \nonumber 
V_{\mathsf{j},\mathsf{j}'}(z)&=\sum_{\mathsf{k, \mathsf{m}}}\mathsf{M}_{\mathsf{j},\mathsf{k}} \Upsilon_{\mathsf{k},\mathsf{m}}(z) \mathsf{M}_{\mathsf{m},\mathsf{j}'}\, e^{i \lambda_{\mathsf{k}} z} \sinh[{r}_{\mathsf{m}}(z)].
\end{align}
This is one of the main results of our contribution: the full diagonalization \D{of the system of Eq. \eqref{one} yields} the general solution in spatio-spectral \DB{fixels} $\mathsf{j}$ (localized, single band) given by Equation (\ref{IndGenSol}). Indeed, the solution for any pump configuration and geometry of the lattice is obtained calculating $\Upsilon_{\mathsf{k},\mathsf{m}}(z)$ and ${r}_{\mathsf{m}}(z)$ from Equation (\ref{intL}). Importantly, the modes \db{$\mathsf{j}$} can be distributed in a quantum network. \DB{In contrast}, the \DB{N-fixel} basis $\mathsf{m}$ (delocalized, broadband) with solution given by Equation (\ref{Csol}) \NBP{is, in general, not directly accessible experimentally as the local oscillator in a homodyne measurement stage should indeed be shaped in a specific spatio-spectral \DB{N-fixel}. This shaping is only possible if spatial and spectral DOF are decoupled. In general, Equation (\ref{intL}) can not be decoupled into a spatial and a spectral part} as the sinus cardinal couples spectral and spatial modes \cite{Kruse2013}. 
Nevertheless, the system \NBP{might} be decoupled in specific cases when the sinus cardinal can be approximated by a separable function like a Gaussian function \cite{Grice2001, Wasilewski2006, Patera2010}. 

\NBP{Equation \eqref{IndGenSol} can be rewritten in terms of CV amplitude and phase quadratures, respectively given by $\hat{x}_{\mathsf{j}}=(\hat{\mathcal{A}}_{\mathsf{j}}+\hat{\mathcal{A}}^{\dagger}_{\mathsf{j}})$ and $\hat{y}_{\mathsf{j}}=i (\hat{\mathcal{A}}_{\mathsf{j}}^{\dagger}-\hat{\mathcal{A}}_{\mathsf{j}})$ (shot noise normalized to 1). This results in a propagator ${\bf{S}}(z)$ for the quadrature vector $\hat{\xi}=(\hat{x}_{1}, \dots, \hat{x}_{\mathsf{N}}, \hat{y}_{1}, \dots, \hat{y}_{\mathsf{N}})^T$, such that $\hat{\xi}(z)={\bf{S}}(z) \,\hat{\xi}(0)$ \cite{Barral2020b}. Gaussian states produced by SPDC are fully characterized by the second-order moments of their quadrature operators. These moments are properly arranged in a covariance matrix ${\bf\Sigma}$ given by 
\begin{equation}\label{cov}
    {\bf\Sigma}(z)={\bf{S}}(z) {\bf 1} {\bf{S}}^{T}(z),
\end{equation}
with ${\bf 1}$ the identity matrix standing for vacuum shot noise. 

Moreover, the effect of propagation losses can be easily included in our model. Considering uniform Gaussian losses for all fixels, we get \cite{Barral2018}
\begin{equation}\label{Loss}
    {\bf\Sigma}_{\eta}(z)=\eta{\bf\Sigma}(z) + (1-\eta) {\bf 1},
\end{equation}
where $\eta \in [0,1]$ is the effective optical transmittance that takes into account both optical losses and efficiency of detection. Typically, the transmittance incorporates the propagation losses, the LO-signal mode overlap, the electronic noise of the homodyne detector, and the quantum efficiencies of the photodiodes \cite{Kaiser2016}. Equation \eqref{Loss} shows how losses couple each mode to a vacuum mode, resulting in an increase of noise --diagonal of ${\bf\Sigma}$-- that degrades the squeezing, and in a decrease of quantum correlations. 
}

\section{Spatio-spectral cluster states}

In the following we illustrate this framework using a relevant example. We analyze the following case: a pump with a flat spatial distribution such that $\eta_{j}=\vert\eta\vert e^{i\phi}$ and $\eta_{k,k'}=\vert\eta\vert \delta_{k,k'} e^{i\phi}$, spectrally Gaussian  
\begin{equation}\nonumber
\Omega(\omega_s + \omega_i) =((2\pi)^{1/2}\Delta_p)^{-1/2} e^{-\frac{(\omega_h-(\omega_s + \omega_i))^{2}}{4\Delta_{p}^{2}}},
\end{equation}
and narrowband --its linewidth (full width half maximum) in intensity $2\sqrt{2\ln(2)}\Delta_p$ much lower than the frexel resolution $\Delta_F$. This pump distribution is coupled in each waveguide producing pairs of photons spectrally symmetric with respect to $\omega_{h}/2$. We set $\vert\eta\vert=1/\sqrt{N}$ and for simplicity we choose $\phi=-\pi/2$ --this is a global phase that will just change the squeezing angle such that $\tilde{\mathcal{L}}_{k,k'} (z)= -i \delta_{k,k'} \,e^{-2i \lambda_{k} z}/\sqrt{N}$. In type 0 (or I) downconversion the wavevector phase-mismatch can be approximated at first order in frequency by $\Delta\beta (\omega_s,\omega_i)\approx \Delta\beta (\omega_h/2,\omega_h/2)+\gamma (\omega_h - \omega_s + \omega_i)$ with $\gamma=(\partial \beta/\partial \omega \vert_{\omega_h} - \partial \beta/\partial \omega \vert_{\omega_(s/i)})$ \cite{Victor2021}. \DB{The phasematch at the degenerate frequency can be achieved by}, for instance, quasi-phasematching and suitable temperature setting. Thus, for a narrowband pump $\Omega(\omega_s + \omega_i)=\delta(\omega_h -\omega_s - \omega_i)$ and $\tilde{\mathcal{L}}^{l,l'} (z)=\delta_{L+1-l,l'}$ in the bandwidth of interest. The joint spatio-spectral distribution is thus 
\begin{equation*}
   \mathcal{L}_{\mathsf{k},\mathsf{k}'}(z)=-i g \sqrt{p_{h}} \delta_{k,k'} \delta_{L+1-l,l'}\,e^{-2i \lambda_{k} z} 
\end{equation*}
where $p_{h}$ is here the pump power per waveguide $p_{h}\equiv P_{h}/N$, such that the spatial modes are decoupled --squeezed-- and the spectral modes are coupled two by two --entangled--. 
Equation (\ref{Hei}) can then be written as
\begin{equation}\label{FPP}
\frac{d\hat{B}_{\mathsf{k}}}{d z}= g  \sqrt{p_{h}}\,e^{-2 i \lambda_{k} z} \hat{B}_{\mathsf{k}'}^{\dag},
\end{equation}
with $\mathsf{k}\equiv \mathsf{k}(k,l)$ and $\mathsf{k}' \equiv \mathsf{k}'(k,L+1-l)$. This is the equation of a two-mode squeezer between \DB{L-fixels} $\mathsf{k}$ and $\mathsf{k}'$ with gain \db{$G\equiv r_{\mathsf{k}}z=[(g  \sqrt{p_{h}})^{2}-\lambda_{k}^{2}]^{1/2}z$.} 
In the individual mode basis, the solution of Equation (\ref{FPP}) for each value of $l$ is given by 
\begin{align} \label{Grid1}
\hat{\mathcal{A}}_{s,j}(z)=\sum_{j'}  (\tilde{U}_{j,j'}(z)\hat{\mathcal{A}}_{s,j'}(0) + \tilde{V}_{j,j'}
(z)\hat{\mathcal{A}}_{i,j'}^{\dagger}(0)),\\ \label{Grid2}
\hat{\mathcal{A}}_{i,j}^{\dagger}(z)=\sum_{j'}  (\tilde{V}_{j,j'}(z)\hat{\mathcal{A}}_{s,j'}(0) + \tilde{U}_{j,j'}^{*}
(z)\hat{\mathcal{A}}_{i,j'}^{\dagger}(0)),
\end{align}
with 
\begin{align}\nonumber
\tilde{U}_{j,j'}(z)&=\sum_{{k}}   {M}_{{j},{k}} {M}_{{k},{j}'}\{\cosh[r_{{k}} z]+i\frac{\lambda_{k}}{ \,r_{{k}}}\sinh[r_{{k}} z ]\}, \\ \nonumber
\tilde{V}_{j,j'}(z)&=\sum_{{k}}   {M}_{{j},{k}} {M}_{{k},{j}'}\,\{\frac{g  \sqrt{p_{h}}}{r_{{k}}}\sinh[r_{{k}} z] \},
\end{align}
and where we have taken $l\equiv s$ (signal) and $L+1-l\equiv i$ (idler). Note that the different shape of Equations \eqref{IndGenSol} and \eqref{Grid1}-\eqref{Grid2} is related to the change of basis back to the \DB{fixel basis from a fully-decoupled N-fixel basis and from a partially-decoupled L-fixel basis, respectively. }

The solution of Equations \eqref{Grid1}-\eqref{Grid2} showing only entanglement between symmetric frexels around the central frequency is the limit of a pump spectrally Gaussian when its linewidth is much lower than the frexel resolution \cite{Ra2025}. A broader pump would include terms modulated by the Gaussian spectral distribution entangling symmetric frexels around $l\pm1$, $l\pm2$, etc; generating correlations mainly between the central frexels with a strength that follows a Gaussian distribution \cite{Kouadou2023, Roman2024}. 

\begin{figure}[t]
\includegraphics[width=0.45\textwidth]{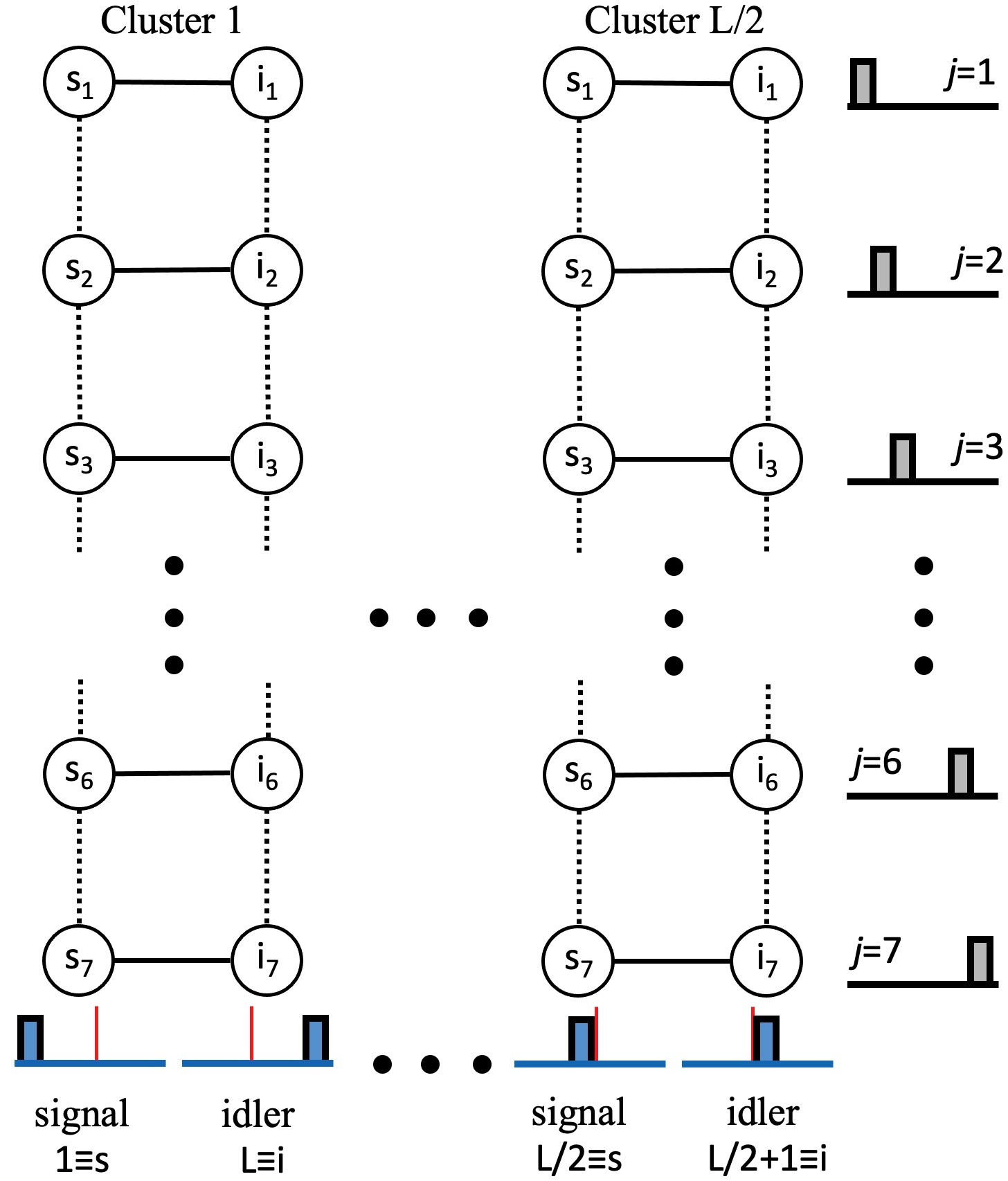}
\caption{\label{F3}\small{\DBR{Set of $7\times 2$-grid cluster states obtained with a monochromatic pump spectrum and a flat spatial pump distribution in a weakly-coupled photonic lattice. Each grid state is composed by $\mathsf{N=}$14 fixels: a pair of frexels $l\equiv s, L+1-l \equiv i$ --symmetric with respect to $\omega_{h}/2$ (in red)-- for each of the seven pixels $j=1, \dots, 7$. Solid lines stand for spectral entanglement (horizontal) and dotted lines stand for spatial entanglement (vertical). $L/2$ ($(L-1)/2$) spectrally-shifted {\it copies} of this state are generated for an even (odd) number of frexels $L$. Each node of the cluster state is in a given spatio-spectral mode representd by gray (spatial) and blue (spectral) rectangles (cf. Fig. \ref{F2}). Blue and black abscissas represent frequency and guided mode, respectively.} }}
\end{figure}

\DB{Remarkably,} in the limit \db{case} of one spectral mode (degeneracy in frequency, $L=1$) the state given by Equations (\ref{Grid1})-(\ref{Grid2}) \db{is} a linear cluster state in the spatial domain over a wide range of values of the governing parameters \cite{Barral2020, Barral2020b}. Hence, the state \db{given by Equations (\ref{Grid1})-(\ref{Grid2})} can form $L/2$ {\it copies} of a \DBR{$(N \times 2)$}-grid cluster state for an even number $L$ of spectral modes, or $(L-1)/2$ copies if $L$ is odd. An example is shown in Figure \ref{F3} for an homogeneous array of nonlinear waveguides with $N=7$ waveguides and two spectral modes $l\equiv s$ and $L+1-l\equiv i$. The horizontal and vertical edges represent respectively spatial and spectral entanglement. For instance, for \DBR{$L=16$} frequency bands we would have \DBR{8} \DB{spectrally-shifted} {\it copies} of the \DBR{$7\times2$} grid state of Figure \ref{F3}. All {\it copies} are generated in the same temporal mode, and can be easily distributable in a quantum network by means of suitable spectral demultiplexing. 
\begin{figure}[h]
     \hspace{0cm}\subfigure{\includegraphics[width=0.41\textwidth]{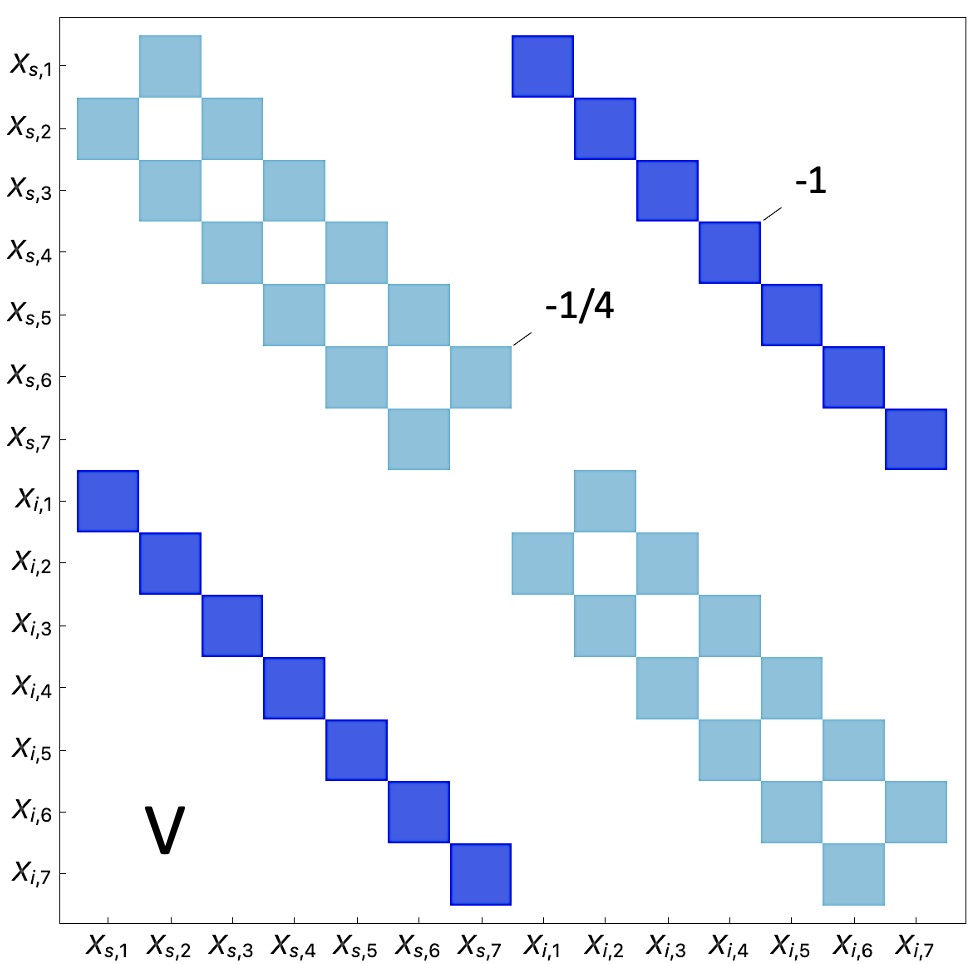}} \\
     \vspace {0.3cm}
    \hspace{0cm}\subfigure{\includegraphics[width=0.46\textwidth]{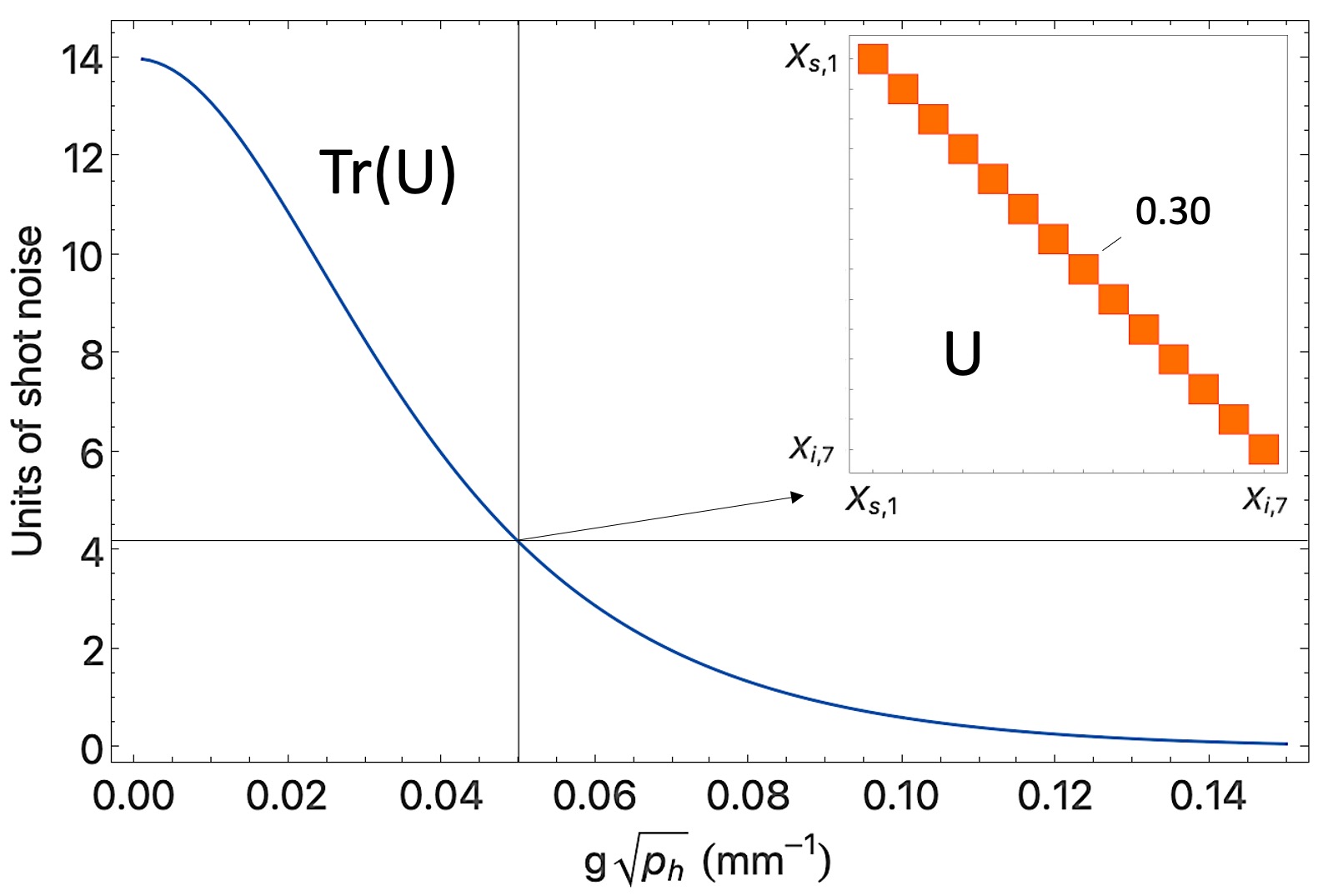}}
\vspace {0.3cm}\,
\hspace{0cm}\caption{\label{F4}\small{Real part ${\bf V}$ (upper figure) of the complex-weighted adjacency matrix ${\bf Z}={\bf V}+i \,{\bf U}$ obtained from Equations (\ref{Grid1}) and (\ref{Grid2}) \NBP{for $L=2$, $N=7$}. ${\bf V}$ is the canonical graph of the state, whereas the trace of the imaginary part ${\bf U}$ accounts for the error of the approximation. The value of Tr(${\bf U}$) for different values of nonlinearity is shown in the lower figure \NBP{(blue curve)}. \DBR{The inset shows the matrix ${\bf U}$ for a specific pump power. As it is diagonal and small, it represents the variances of the graph nullifiers. $x_{s(i),j}$ is the amplitude quadrature of a fixel $\mathsf{j}$ (pixel $j$, frexel $l$).}
The vacuum shot noise is set as 1. We have applied a $\pi/2$ rotation in idler-mode phase space (exchange of labels of quadratures for the idler modes), and used a homogeneous coupling profile $\vec{f}=\vec{1}$ with $C_{M}=0.01$ mm$^{-1}$, $g \sqrt{p_{h}}=0.05$ mm$^{-1}$ (upper figure, \DBR{lower figure  inset}) and $z=20$ mm.
}}
\end{figure}

\DBR{We can demonstrate that the array produces this family of states using the graph calculus for Gaussian pure states \cite{Menicucci2011}. Using Equations (\ref{Grid1}) and (\ref{Grid2}), we  can calculate the covariance matrix ${\bf\Sigma}$ given by Equation \eqref{cov}, but in this case related to the quadrature vector $\hat{\xi}=(\hat{x}_{s,1}, \hat{x}_{s,2}, \dots, \hat{x}_{i,1}, \hat{x}_{i,2}, \dots, \hat{y}_{s,1}, \hat{y}_{s,2}, \dots, \hat{y}_{i,1}, \hat{y}_{i,2})^T$. Notably, ${\bf\Sigma}$ can be written in block form in terms of two symmetric matrices ${\bf U}$ and ${\bf V}$ as \cite{Menicucci2011}
\begin{equation*}
    {\bf\Sigma}(z)=\begin{pmatrix}
        {\bf U}^{-1} & {\bf U}^{-1} {\bf V}  \\ {\bf V}{\bf U}^{-1}  & {\bf U}+{\bf V}{\bf U}^{-1}{\bf V}
    \end{pmatrix}.
\end{equation*}
 From the two upper blocks of ${\bf\Sigma}$ we can obtain the complex-weighted adjacency matrix ${\bf Z}={\bf V}+i \,{\bf U}$ that fully characterizes Gaussian pure states \cite{Menicucci2011}. The real part ${\bf V}$ is the graph of the ideal CV cluster state approximated by ${\bf Z}$. Its  nullifiers are $\bf{\hat{\delta}}=\bf{\hat{y}}-V \bf{\hat{x}}$, where $\bf{\hat{x}}$ and $\bf{\hat{y}}$ are vectors of amplitude and phase quadratures, respectively. If the imaginary part ${\bf U}$ is small, it can be interpreted as the covariance matrix of the nullifiers $\bf{\hat{\delta}}$. The elements in its diagonal Var($\hat{\delta}_j)={\bf U}_{j,j}$ thus represent the variances of the nullifiers, with its trace Tr(${\bf U}$) being the error of approximating the ideal cluster state ${\bf V}$ with ${\bf Z}$ \cite{Menicucci2011}. Figure \ref{F4} shows an example of the real (upper) part and the trace of the imaginary part (lower) of ${\bf Z}$ obtained using Equations (\ref{Grid1}) and (\ref{Grid2}) for $N=7$ and $L=2$. We obtain a non-unit weight matrix ${\bf V}$ with negative entries and the weight of the spectral vertices {$(s_{j}:i_{j})$} four times that of the spatial vertices {($s_{j}:s_{j\pm1}$), ($i_{j}:i_{j\pm1}$)}  (Figure \ref{F4} upper). For the parameters used in our simulation there are also negligible spatio-spectral correlations {($s_{j}:i_{j\pm1}$) and ($i_{j}:s_{j\pm1}$)} more than ten times lower than the spatial ones {(not shown for the sake of clarity)}. From this ${\bf V}$, we can write down an approximation of the nullifiers of the state
\begin{align}\label{Nul1}
\hat{\delta}_{s,j} &= \hat{y}_{s,j} + \frac{1}{4}(\hat{x}_{s,j-1}+\hat{x}_{s,j+1}) + \hat{y}_{i,j}, \\ \label{Nul2}
\hat{\delta}_{i,j} &= -\hat{x}_{i,j} + \frac{1}{4}(\hat{y}_{i,j-1}+\hat{y}_{i,j+1}) + \hat{x}_{s,j},
\end{align}
for $j=1,\dots, 7$, and with a $\pi/2$ rotation in the idler-mode phase space. The inset in Figure \ref{F4} (lower) shows the matrix ${\bf U}$, that can be interpreted as the covariance matrix of the nullifiers for ${\bf U}$ small (here $g \sqrt{p_{h}}$ large). Here, the variances are approximately equal and squeezed --well below their respective shot noise level for both 3- and 4-node nullifiers, 33/16 and 17/8, respectively.} Hence, for a given length of the array, in the limit of weak evanescent coupling and large nonlinearity, the state given by Equations (\ref{Grid1}) and (\ref{Grid2}) resembles the grid state of Figure \ref{F3} with the error vanishing for infinite squeezing (Figure \ref{F4} lower). This result is expected since for \DB{weak evanescent coupling SPDC light produced in each waveguide {\it spreads} only into} nearest-neighbor waveguides, but not beyond. In the limit of high coupling the state will present a different entanglement geometry \cite{Barral2021PRR}. \DBR{Note that this analysis is applicable to any number of spatial and spectral modes, $N$ and $L$, respectively.} 


\section{Feasibility and outlook}

\begin{figure}[t]
     \includegraphics[width=0.48\textwidth]{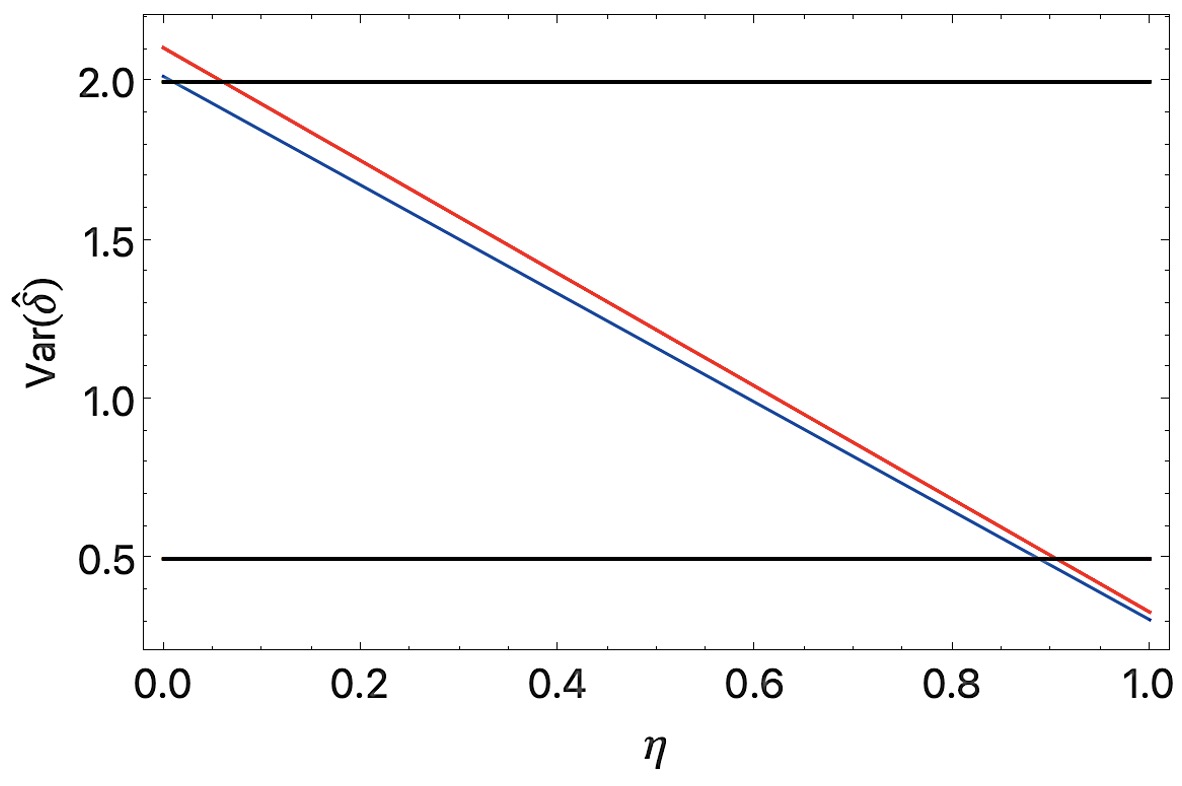}
    \caption{\label{F5}\small{\DBR{Effect of losses on the variances of the nullifiers. Dependence of 3-node and 4-node variances with $\eta$ are in blue and red, respectively. Black horizontal lines represent the squeezing threshold per nullifier for spatial (Var$(\hat{\delta})<1/2$) and spectral (Var$(\hat{\delta})<2$) full inseparability.}}}
\end{figure}

\NBP{We analyze now the effect of losses on the generation and measurement of the grid cluster state. The analysis of the previous section only applies for Gaussian pure states, so we cannot apply it in the present case. What we can do however is to calculate the variance of the nullifier given by Equations (\ref{Nul1})-(\ref{Nul2}) directly from the elements of a lossy covariance matrix as in Equation \eqref{Loss}. In Figure \ref{F5}, we show how Var($\hat{\delta})$ varies with respect to an effective optical transmittance $\eta$ that characterizes uniform Gaussian losses. The first figure of merit is nullifier squeezing. The calculation shows that 3- and 4-node nullifiers, in blue and red, respectively, are squeezed while $\eta \neq 0$. A second figure of merit is nullifier entanglement, usually certified experimentally by the violation of van Loock-Furusawa inequalities \cite{VLF2003}. Here we get two bounds of full separability, one for spectral entanglement and the other for spatial entanglement. They are given respectively by
\begin{align}\label{VLF1}
{\text Var}(\hat{\delta}_{s,j}) &+ {\text Var}(\hat{\delta}_{i,j}) \geq 4, \\ \label{VLF2}
{\text Var}(\hat{\delta}_{s(i),j}) &+ {\text Var}(\hat{\delta}_{s(i),j+1}) \geq 1.
\end{align}
Thus, spectral entanglement appears between any pair of nodes minimally squeezed (Var($\hat{\delta})<2$) whereas spatial entanglement requires a large squeezing (Var($\hat{\delta})<1/2$) to be certified. These bounds are displayed in Figure 5, showing that spatial entanglement is lost above $\approx10\%$ losses. This value is challenging but feasible, as detection optical losses as low as 8\% have been recently reported \cite{Kashiwazaki2023}. Interestingly, this threshold remains independent of the total number of spatio-spectral modes, ensuring that scalability is not affected. Moreover, this limitation can be alleviated by building up the initial squeezing by increasing the value of $g \sqrt{p_{h}}$. Note that this is a consequence of the weak coupling regime set in this specific example. Other regimes with larger coupling --producing a different type of cluster states-- will not suffer from this issue. 
}

\NBP{An optical phase locking system is essential to distribute the cluster nodes in a quantum communication network while keeping the relative phases between nodes constant. This can be accomplished using a classical frequency comb that is phase-locked to the broadband SPDC: classical-comb frequency bands will act as the dedicated pilot tone for their corresponding SPDC spectral band. Typically, a mode-locked laser could act as master laser, being split into a LO, an auxiliary (seed) field, and a field to be upconverted by second harmonic generation to generate the pump \cite{Roslund2014, Cai2017, Kouadou2023, Roman2024}. In a hold-and-measure optical phase locking, the seed takes the role of the SPDC field along a chopping cycle, allowing one to retrieve phase information from the interaction of the seed with the pump and LO fields \cite{Roeland2023}. Since pump and SPDC phases are locked by the nonlinear interaction, setting the relative phases between seed and pump and seed and LO through suitable detection and phase feedback, enables locking of the LO-SPDC phase over time. Once the seed is blocked, the feedback loop keeps the phase stable for a period longer than the measurement time, time used to measure the correct quadrature at each node.
}



Grid cluster states are an important resource for measurement-based quantum computing \cite{Menicucci2006}. \NBP{The limit in the size of the grid of Figure \ref{F3} depends mainly on the number of pixels --waveguides-- in the lattice. Arrays with tens of modes ($N$) have been recently demonstrated \cite{Yang2024, Raymond2024}. In addition, shaping spatially the pump has been demonstrated in different substrates \cite{Jin2014, Lenzini2018, Barral2021OE, Bao2023}. The low efficiency of the $\chi^{(2)}$ nonlinearity in waveguides could nonetheless hamper the scalability of the system due to the need of large amounts of pump power per waveguide. This issue can be alleviated with thin film waveguides with efficiencies twenty times that of standard waveguides \cite{Wang2018,Stokowski2023}. Notably, using an efficiency for second harmonic generation of 2600\% W$^{-1}$ cm$^{-2}$ as reported in \cite{Wang2018}, only 10 mW of pump power per waveguide would be necessary to reach the value $g \sqrt{p_{h}}=0.05$ mm$^{-1}$ used for the simulation of Figure \ref{F4}. On the spectral side, the number of frexels depends mainly on the spectral width of the local oscillator ---that can be coherently broadened using an all normal dispersion (ANDi) photonic crystal fiber producing low-noise self-phase modulation \cite{Renault2023}-- and the dispersive power of the diffractive element used to demultiplex frequencies \cite{Michel2021}. Notably, homodyne detection with up to $L=16$ frexels have been demonstrated over bandwidths of tens of $nm$ \cite{Roslund2014, Cai2017, Kouadou2023, Roman2024}. Taking all the above into account, achieving graph states with $\mathsf{N}=N \times L\approx 100$ spatio-spectral modes seems possible with current technology. }

Larger grid cluster states can be created using temporal modes under suitable temporal multiplexing \cite{Larsen2019, Asavanant2019}. For instance, we can obtain grid states composed of \DBR{$N\times L$} elements by multiplexing linear clusters in frequency and space encodings. This state is generated at the repetition rate of the pump laser and can be time-multiplexed by applying delay lines. Thus, the ability to shape the pump field in both space and frequency in a nonlinear photonic lattice opens a wide range of possibilities to create two- and three-dimensional cluster states, instrumental respectively for universal and fault-tolerant measurement-based quantum computing \cite{Bourassa2021}. For instance, suitable spatial pump shaping can produce a closed linear cluster that multiplexed in frequency results in a torus in a single temporal mode \cite{Barral2020}.

Another option of practical interest is to use individual modes in the spatial domain and squeezed supermodes in the frequency domain with a flat spatial distribution of the pump fields. In this case there is a number of independently squeezed spectral supermodes for each $j$ spatial mode \cite{Fabre2020}. For instance for both Gaussian phasematching and pump spectrum, the spectral supermodes are close to Hermite-Gaussian modes \cite{Patera2010}. This \db{spectral} basis is \DB{reachable} through suitable LO pulse shaping in homodyne detection. Notably, in this case \db{the independently-squeezed Hermite-Gauss spectral supermodes are} entangled due to spatial correlations. \D{Thus, in the example above we would obtain $L$ {\it copies} of a $N$-mode linear cluster state}, each one with a different Hermite-Gaussian spectrum. 

\DB{De-Gaussifying quantum states generated in the optical lattice can be easily accomplished by photon subtraction in the fixel basis using just a dispersive element that isolates a specific fixel and a high-transmission beam splitter. This is in contrast with broadband-frequency modes that require a nonlinear interaction with a gate beam that selects the spectral mode of subtraction \cite{Ra2020}. Fixel-based photon subtraction avoids crosstalks natively as is carried out in a frequency band occupied by just one spectral mode, unlike spectrally-broadband photon subtraction where each mode occupies simultaneously a number of bands increasing the risk of crosstalk and thus of purity degradation. The non-Gaussianity then spreads over the cluster opening the possibility to engineer multimode non-Gaussian states \cite{Walschaers2017},} \D{a crucial resource for quantum advantage \cite{Chabaud2023}}.

\NBP{
In general, CV graph states of any size can be generated in the spatial domain via individual squeezers and universal optical circuits like a mesh of Mach-Zehnder interferometers \cite{Asavanant2024}. In contrast, our system gives access to a limited subset of all possible graph states. What subset is is an open question. As an outlook, a group theory analysis would determine the set of graph states that is accessible through pump shaping in nonlinear optical lattices. However, the subset accessible via nonlinear optical lattices have a better scaling than optical linear circuits, as the size of the array will be similar to that of the individual waveguide squeezers section, saving the space corresponding to the interferometer section. Moreover, the tunability in the state generation is in our case inherited from the classical-pump pulse shaping in contrast to the linear optical circuit which acts directly on the generated quantum states, that are more sensitive to propagation losses. 
}

Finally, we note that in general type-0 and type-I SPDC in few-cm waveguides with periodic poling like PPKTP and PPLN present a broad $\Delta_{SPDC}$ limiting the squeezing available per frequency band \cite{Roman2024}. Group velocity matching in type-II and, notably, in type-I SPDC has been recently proposed as a possible solution for large squeezing in a limited bandwidth in cm-scale waveguides \cite{Victor2021, Horoshko2025}. \DBR{Moreover, poling engineering in $\chi^{(2)}$ nonlinear waveguides enables the generation of tailored spatio-spectral modes that can improve resources like squeezing or entanglement \cite{Barral2019,Weiss2025}.}

\section{Conclusions}

As a summary, we have analyzed the spatio-spectral features of a $\chi^{(2)}$ nonlinear photonic lattice, presented the general solution of the system in the low gain regime, and discussed the possibilities to generate large and distributable two-dimensional \DB{spatio-spectral} cluster states with a single integrated source through suitable spatio-spectral pump shaping. Recent demonstrations of squeezing in single nonlinear waveguides in spatial and spectral modes \cite{Nehra2022, Kashiwazaki2023, Kouadou2023, Roman2024}, and of second harmonic generation in arrays of nonlinear waveguides and in slab waveguides with arbitrarily reconfigurable two-dimensional distribution of nonlinearity \cite{Barral2021OE,Yanagimoto2025}, open the door to realizing fully integrated multimode spatio-spectral squeezing in optical lattices, paving the way for scalable quantum photonic technologies as quantum networks, distributed quantum sensing and universal and fault-tolerant measurement-based quantum computing.

\section*{Acknowledgements}

This work was supported by MICINN through the European Union NextGenerationEU recovery plan (PRTR-C17.I1), the Galician Regional Government through “Planes Complementarios de I+D+I con las Comunidades Autónomas” in Quantum Communication, and the Paris Ile-de-France region in the framework of DIM SIRTEQ.

\section*{Appendix}
\renewcommand{\theequation}{A.\arabic{equation}}
\setcounter{equation}{0}

We analyze below the approximation over which our model \DB{is valid}. For that, we describe the dependence of the coupling with the wavelength following the model of \cite{Kruse2013}, but making explicit the dependence with the distance between waveguides as in \cite{Barral2021OE}. We write thus
\begin{equation}\label{A1}
C(\lambda,d)=\frac{C_{0}}{\lambda} e^{-\Gamma_{0} \frac{n(\lambda)}{\lambda}d },
\end{equation}
where $\lambda$ is the wavelength, $n(\lambda)$ the refractive index in the waveguides, $d$ the distance between waveguides, and $C_{0}$ and $\Gamma_{0}$ are constants. This model is suited for distances $d$ larger than a minimal distance where next-to-nearest neighbor effects are  negligible. For instance, using the data of \cite{Barral2021OE} for a nonlinear directional coupler in LN, we get $C_{0}=25.6$ $\mu m/ mm$ and $\Gamma_{0}=0.19$ for $d \geq 13$ $\mu m$ and $\lambda$ in $\mu m$. 

In order to analyze the dependence of the coupling profile with the wavelength we need firstly to define it. The coupling profile is mapped to a set of distances between waveguides --or distance profile. Experimentally, we define a set of distances that map the coupling at a given wavelength $\lambda_0$ using Equation \eqref{A1}. The largest coupling of the profile $C_M$ will have associated the shortest interdistance $d_{m}$ as
\begin{equation}\label{A2}
C_{M}=\frac{C_{0}}{\lambda_0} e^{-\Gamma_{0} \frac{n(\lambda_0)}{\lambda_0} d_{m} }.
\end{equation}
We calculate the coupling between waveguide $j$ and $j+1$ writing Equation \eqref{A1} in terms of $C_M$ and the $j$th interdistance $d_j$
\begin{equation}\label{A3}
C_{j}(\lambda,d_j)=C_{M}\frac{\lambda_{0}}{\lambda} e^{-\Gamma_{0}(\frac{n(\lambda)}{\lambda}d_{j}-\frac{n(\lambda_{0})}{\lambda_{0}}d_{m}) }\equiv C_{M} f_{j}(\lambda).
\end{equation}
This equation defines a coupling profile wavelength dependent $\vec{f}(\lambda)$. The (normalized) coupling profile of design is however $\vec{f}^{D}=\vec{f}(\lambda_0)$ with $f_j^D \in [0,1]$. For example, for a Glauber-Fock array the normalized coupling profile is $f_{j}^{D}=\sqrt{j/(N-1)}$ for $j=1,\dots, N-1$ \cite{Barral2020b}. The experimental set of distances corresponding to this coupling profile is 
\begin{equation}\label{A4}
d_{j}=d_{m} - \frac{\lambda_0}{\Gamma_{0} n(\lambda_{0})} \ln(f_j^D).
\end{equation}

The coupling profile is distorted at wavelengths $\lambda$ off $\lambda_{0}$ and given by
\begin{equation*}
   f_{j}(\lambda)=\frac{\lambda_0}{\lambda} e^{\frac{\lambda_{0}}{\lambda}\frac{n(\lambda)}{n(\lambda_{0})} \ln(f_{j}^D)} e^{-\Gamma_{0} (\frac{n(\lambda)}{\lambda}-\frac{n(\lambda_0)}{\lambda_0})d_{m}}.
\end{equation*}
The above equation shows that the profile distortion is more evident as the distance to the design wavelength $|\lambda-\lambda_{0}|$ increases and $f_{j}^{D}$ decreases. For instance, for an array in LN with $d_0$=13 $\mu m$, $\Gamma_{0}=0.19$, and coupling profile elements at the design wavelength $f_{j}^{D}(1.55\, \mu m)=1(0.1)$, we get a slight increment of up to $\pm 5(10)\%$ in a bandwidth of 60 $nm$ and up to $\pm 8(16)\%$ in 100 $nm$. This variation is lower than that produced by fabrication errors. In general, the distortion of the coupling profile --and thus of the supermodes-- for this array would be negligible over a bandwidth above $60$ $nm$ and, for the particular case of an homogeneous coupling profile where $f_{j}^{D}(1.55\, \mu m)=1$ for all $j$, there would be no perceptible effect over a bandwidth above $100$ $nm$. For larger/shorter wavelengths the supermodes would \D{retain its shape, but would experience} different effective propagation lengths.


\end{document}